# Non-Destructive Rail Monitoring for Defect Identification

Elissa Akiki [1,2], Lynda Chehami[1], Nikolay Smagin[1], Emmanuel Moulin[1], Jamal Assaad[1], Youssef Zaatar[2]

[1] Polytechnic University of Hauts-de-France, IEMN – Institute of Electronics, Microelectronics and Nanotechnology, Valenciennes site, CNRS UMR 8520, Le Mont Houy, 59313 Valenciennes Cedex 9, France

[2] Lebanese University – Faculty of Sciences II, Pierre Gemayel Campus – Fanar, P.O. Box 90656 Jdeidet, Lebanon

**ABSTRACT**

*Non-destructive evaluation (NDE) of rail tracks is crucial to ensure the safety and reliability of rail transportation systems. In this work, we present a quantitative study using various signal processing methods to identify defects in rail structures. A diffuse field configuration was employed at few dozens of kiloHertz, where the emitter and receiver were remotely located, and wave energy propagated via multiple reflections within the medium. A reference database is first constructed by acquiring measurements at different rail positions and different torque levels (up to 50 N.m). The defect is then identified by comparing its signature to those stacked in the database. First, the destretching technique, based on Coda Wave Interferometry (CWI), is applied to correct for temperature-induced velocity variations. Then, the identification is performed using the Mean Square Error (MSE) metric and Orthogonal Matching Pursuit (OMP) technique. A comparative analysis of the both methods is conducted, focusing on their robustness and performance.*

## 1. INTRODUCTION

Monitoring and controlling railway systems remains a major challenge across industries, particularly for rail tracks. These components are exposed to frequent service loads and harsh environmental conditions [1,2]. When subjected to dynamic stresses and aggressive environmental impacts, railway tracks are prone to developing internal cracks and other flaws. These defects can give rise to unwanted vibrations and noise even at an early stage, potentially reducing acoustic comfort, degrading system performance, and—if left unaddressed—compromising structural integrity and safety over time. Addressing these issues is a critical priority for both researchers and engineers.

Different signal processing techniques based on optimization problems are proposed, such as Mean Squared Error [3], used, for example, in passive structural health monitoring [4,5]. Among other advanced approaches, orthogonal matching pursuit (OMP) methods offer an alternative technique by isolating meaningful components through optimized projections, effectively filtering out noise and emphasizing relevant features [6].

Orthogonal Matching Pursuit (OMP) is a widely used sparse approximation method for signal reconstruction and variable selection. It is a greedy algorithm that aims to represent a given signal as a linear combination of a small number of windows (called "atoms") from an overcomplete dictionary. At each iteration, OMP selects the atom that is most correlated with the current residual and updates the solution by minimizing the reconstruction error. This strategy provides a sparse solution with fast convergence under suitable conditions. Due to its simplicity and efficiency, OMP is commonly applied in fields such as signal processing, machine learning, and image analysis, where sparse data representations are crucial [6,7].

Da-Zhi Dang et al. [8] proposed, for example, a guided wave testing (GWT) method using optical fiber sensing combined with an orthogonal matching pursuit (OMP)-based data processing technique to detect rail damage by reconstructing defect-related reflective waves from complex raw signals. Similarly, Zhang et al. [9] applied OMP to guided wave signal analysis to improve defect detection in metallic structures.

Despite their strengths, these methods require, a priori, compensation for environmental conditions (such as temperature). For this, coda wave interferometry (CWI) offers an elegant way to suppress or compensate for temperature bias.

CWI was initially introduced by Snieder et al. [10] as a seismic method to evaluate local variations in wave velocities in the ground. The coda signal consists of waves reflected multiple times off the structure boundaries and heterogeneities. This signal is very reproducible as long as no changes occur in the elastic properties or geometry of the tested structure. However, if any alternation of propagation media appears, small delays are introduced on the wave paths and will accumulate over multiple reflections. As a result, the coda signal is sensitive to damage that would remain undetectable by first arrival waves. The principle of CWI is to compare, e.g., with correlation, the state of the coda at a given time with that of a reference coda signal measured for the structure's initial state.

Wang et al. [11] showed that CWI can track real-time local variations of wave velocity in complex media, enabling continuous monitoring of structural changes. It is worth noting that the coda signal is not only sensitive to the appearance of damage but is also extremely sensitive to any environmental changes that affect wave velocities in the structure, such as temperature variations [12]. The objective of this research work is to use the destretching procedure common to the CWI as a tool for temperature compensation and then apply MSE and OMP techniques for defect identification in rail structure. To the authors' best knowledge, CWI combined with OMP has never been used to monitor rail tracks

This paper is organized as follows: the experimental setup and associated signal processing are presented in Section 2. Section 3 is devoted to the obtained results and discussion. Finally, conclusions are drawn from the different obtained results.

## 2. ULTRASONIC MEASUREMENTS AND ASSOCIATED SIGNAL PROCESSING

Ultrasonic measurements were conducted on a two-meter-long rail equipped with two piezoelectric sensors (one referred as emitter "E" and the other as receiver "R") (see Figure. 1). This configuration enables precise measurement of the acoustic signals propagating along the rail, allowing for the characterization of the structure's acoustic properties, such as propagation speed and signal amplitude.

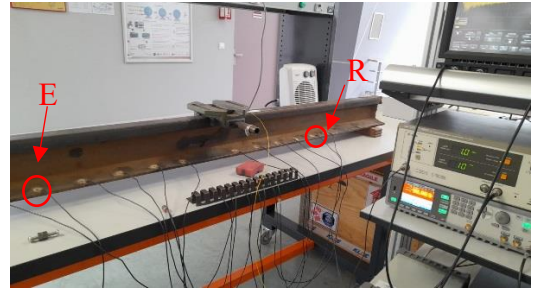

Figure 1: Experimental setup

As part of the experimental setup, the "artificial" defect was simulated using a vise fixed to the rail head, where the three torque levels were applied using a dynamometric torque wrench (see Figure 1).

Measurements were recorded at 20 locations spaced by 10 cm intervals using torque levels of 10, 25, and 50 N.m. To generate broadband coda waves, a linear chirp signal was applied on the transducer, featuring a frequency sweep from 10 kHz to 500 kHz, with a duration of 1 ms.

The reference dictionary is used to train the defect detection model, which is constructed by measuring acoustic signals at various positions along the rail under three levels of applied torque (10, 25, and 50 N.m). To eliminate environment variability, the signals are preprocessed by correcting them using Coda Wave Interferometry (CWI), and then they are averaged at each position to generate a representative reference signal.

### 2.1 CWI Principle

Coda Wave Interferometry (CWI) relies on the difference between the coda waves observed before and after a propagation medium. These waves, resulting from multiple reflections in the medium, are very sensitive to tiny velocity or structural changes [13].

The principle assumes that the disturbed signal is a time-stretched copy of the reference signal. That is, a velocity change (δV/V) in the medium causes the waveform to expand or collapse in time. The stretch is measured using the stretched correlation coefficient denoted by "$\alpha$".

First, the correlation coefficient "$R$" is computed between both perturbed "$u_{per}$" and the reference "$u_{ref}$" signals:

$$R(t_1, t_2)(\varepsilon) = \frac{\int_{t_1}^{t_2} u_{ref}(t) \cdot u_{per}(t[1+\varepsilon]) \, dt}{\sqrt{\int_{t_1}^{t_2} u_{ref}^2(t) \, dt \cdot \int_{t_1}^{t_2} u_{per}^2(t[1+\varepsilon]) \, dt}} \quad (1)$$

where:
- $\varepsilon$ is the extension factor,
- $[t_1, t_2]$ is the time window in the coda.

       

The ε value that maximizes the coefficient $R$ provides relative velocity change $\frac{\delta V}{V} = -\varepsilon$, which allows precise measurement of changes in the medium.

Thus, the stretched correlation coefficient $\alpha$ is defined as:

$$\alpha = \mathrm{argmax}(R(t_1, t_2)(\varepsilon)) \quad (2)$$

The signals are therefore re-stretched by this value $\alpha$ in order to compensate for the temperature effect.

The perturbed signals are then destretched by this coefficient $\alpha$, in order to correct for the temperature induced time dilation, hence we got $u_{dest}$ :

$$u_{dest} = u_{per}(t[1 - \alpha]) \quad (3)$$

The time delay is then determined using cross-correlation between the signals, followed by a polynomial fit of the correlation function. This involves approximating the shape of the correlation peak with a parabolic curve around its maximum (or minimum). This fitting allows for a precise estimation of the time delay, especially when the sampling frequency is insufficient to resolve sub-sample time shifts directly. Such a refinement is crucial for accurate temporal alignment between signals.

To analyze the effect of temperature on wave velocity, various stretching factors ranging from -10% to 10%, were applied to the temperature-related signals. These stretching factors represent relative velocity changes between the temperature signals. By adjusting the cross-correlation between signals using these different stretching values, the impact of these variations on temperature measurements can be analyzed. This process mimics the effect of thermal expansion or contraction of the medium, which can lead to apparent velocity changes in wave propagation.

The reference dictionary was constructed by averaging all signals at each time step after CWI has been performed, thereby enhancing the signal-to-noise ratio and minimizing temperature variations. Destretching was similarly employed as a preprocessing technique to each blind test signal for homogenizing preprocessing for all datasets.

The coda waves, which correspond to the late-arriving scattered waves in the signal, are particularly sensitive to subtle changes in the medium's properties, making them ideal for detecting small velocity variations (see Figure 2).

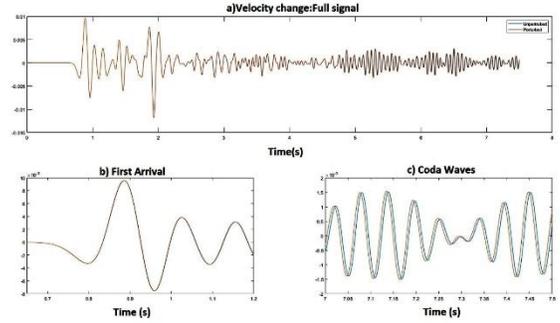

Figure 2: The complete signal, the first incoming waves and the coda waves.

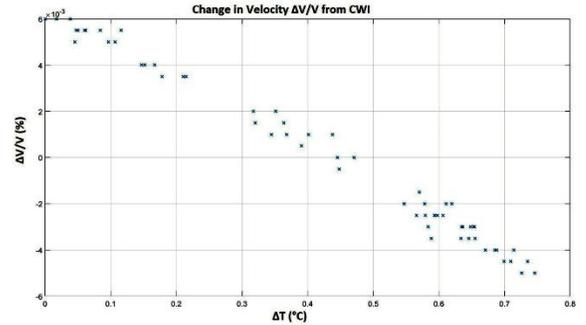

Figure 3: ΔV/V vs. Temperature Change (ΔT)

Figure 3 illustrates the relative change in wave velocity (ΔV/V, expressed in %) as a function of the temperature variation (ΔT, in °C). Each point represents a single measurement where the temperature has changed relative to the reference condition.

As expected, the time delay evolves linearly with time, reflecting the cumulative effect of the relative velocity change over the entire signal window. This linear trend confirms the theoretical expectation that uniform velocity perturbations manifest as linear drifts in the arrival times of coda waves. The slope of this curve is directly related to the relative change in wave velocity, thus offering a quantitative measure of the thermal influence on the medium.

This underlines the importance of the destretching process, which is carried out by taking into account the value of ΔT (the time shift) used to adjust the signals accordingly [10]. Once ΔT has been estimated, signal interpolation is performed, and the signals are then restretched using these values to apply the necessary correction.

**2.2 Mean Squared Error**

Let us define $S_{ij}$ as the signals measured at receiver $R$ (see Figure.1) for each position '$j$' and torque level '$i$'. As mentioned before, the reference training database ($D$) is the average over $N$ torque levels; i.e.;

$$D_j = \frac{1}{N}\sum_{i=1}^{N} S_{ij}(t) \quad (4)$$

where **N** corresponds to the number of applied torque levels.

The Mean Squared Error (MSE), which quantifies the difference between the actual and predicted signals denoted by $S_{test}$, is used to measure the average squared deviation between the test signals and the reference signals from D. In our case, the MSE is computed for each position j between the test signals and the reference dictionary signals, as expressed by the following equation:

$$MSE_j = \int [S_{test}(t) - D_j(t)]^2 dt \quad (5)$$

The integration is performed over the considered time range.

For more clarity, we define a damage identification index as the inverse of the MSE denoted by $r_j$,

$$r_j = \frac{1}{MSE_j} \quad (6)$$

This parameter provides a quantitative measure of identification accuracy. A lower MSE results in a higher identification score, indicating a better match between the test signal and the reference signal.

## 2.3 Orthogonal Matching Pursuit (OMP)

The OMP algorithm seeks a sparse approximation of a test signal $S_{test}$ using a dictionary D composed of reference signals.
The main objective is to minimize the residual between the original signal and its approximated using the fewest number of dictionary elements (atoms). In so doing, it is expected that the test signal is efficiently represented by a limited collection of representative signals (atoms) in the dictionary, adhering to the principle of parsimony or sparsity.
In our case, each atom in the dictionary represents a signal from a given position and force on the rail. The atom selected with the highest coefficient is therefore the one whose profile best matches the test signal, i.e., the estimated position of the fault.

Mathematical Formulation:
The signal $S \in R^N$ is approximated as a sparse linear combination of atoms from the dictionary:

$$S \approx D_x = \sum_{i \in S} x_i d_i \quad (7)$$

Where:
- $D = [d_1, d_2, \ldots d_M] \in R^{N*M}$ is the dictionary matrix
- $x \in R^M$ is the sparse coefficient vector with few nonzero entries
- S is the active support set, i.e., indices of selected atoms

**Algorithm steps**
Initialization:
- Residual $r^{(0)} = S$
- Support set $S^{(0)} = \emptyset$
- Iteration counter k=0

For k=1…,$K_{max}$, do:

✓ Compute correlations between the current residual and all dictionary atoms: $c = D^T r^{(k-1)}$ (8)

✓ Apply threshold τ on correlations

✓ Select the atom $i_k$ with the maximum absolute correlation: $i_k = \text{argmax}|c_i|$ (9)

✓ Find best approximation coefficients on the selected support: $S^{(k)} = S^{(k-1)} \cup \{i_k\}$ (10)

✓ Find the best approximation coefficients $x^{(k)}$ on the selected support by solving a least squares problem:
$$x^{(k)} = \text{argmin} \left\| S - D_{S^{(k)}} x \right\|_2^2 \quad (11)$$

Where:
$D_{S^{(k)}}$ is the sub-dictionary composed of atoms indexed by $S^{(k)}$.

✓ Update the residual: $r^{(k)} = S - D_{S^{(k)}} x^{(k)}$ (12)

Stop if the residual norm $||r_k||_2$ is sufficiently small or if k= $K_{max}$.
The detected position corresponds to the dictionary atom with the maximum amplitude in the coefficient vector x.

## 3. RESULTS AND DISCUSSION

Here we report all the obtained results for the both methods applied after CWI-based temperature compensation. First, we consider a situation when the considered blind test signals exist in the dictionary.

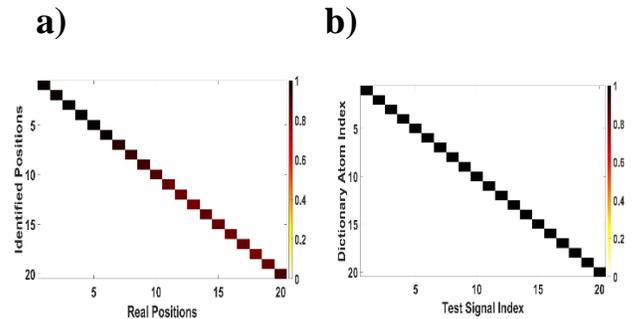

a)          b)

Figure 5: Obtained results: (a) with MSE (b) with OMP, both applied to the average response across the three torques at each dictionary position.

The figures show the Mean Squared Error and OMP values calculated between the measured signals and the averaged reference signals for each tested defect position. Each row corresponds to a different torque level (10 N.m, 25 N.m, 50 N.m), while each column represents a specific defect position along the rail.

As can be observed, the defect identification is accurately estimated based on the average response across the three torque levels. Similar accuracy is obtained when considering each individual torque level.

Now, to test the robustness of the methods, the same procedure was applied to more challenging cases, i.e., signals outside the dictionary taken at arbitrary positions between the emitter (E) and receiver (R). The results are shown in Figure 6.

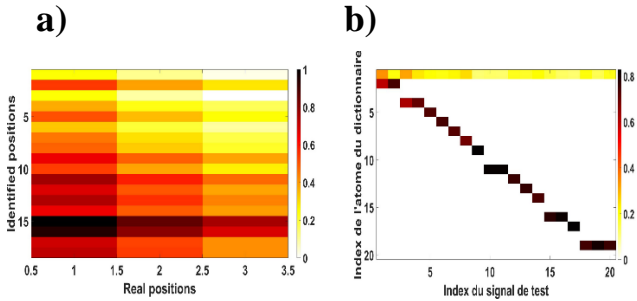

Figure 6: Obtained results: a) with MSE, b) with OMP both applied to the average response across the three torques at each dictionary position

The results demonstrate that the MSE method does not allow for effective defect identification, whereas OMP provides better results and enables correct localization.

## 4. CONCLUSION

This paper focuses on defect identification in rail track structures using advanced signal processing techniques. First, ultrasonic measurements at a few kHz are performed, and Coda Wave Interferometry (CWI) is applied to compensate for temperature bias. Then, MSE and OMP methods are used to compare blind test signals to a training database in two scenarios: when the tested signals are inside the database and when they are outside. The obtained results are satisfactory, and the comparative study shows that OMP is very robust, especially in the challenging case where the defect is outside the database. These results represent a first step toward passive rail tracks monitoring where acoustic signal excitation will be replaced by natural ambient noise.

**ACKNOWLEDGEMENTS**

This work is funded by the Hauts-de-France region and the company Transmotor in Lebanon. This work was partly funded by the French ANR agency: DACLOS: ANR-21-CE42-0002.